# Second-Guessing in Tracing Tasks Considered Harmful?

Bhushan Chitre[1], Jane Huffman Hayes[1], and Alexander Dekhtyar[2]

[1] University of Kentucky, Computer Science, Lexington, Kentucky, USA
[2] California Polytehcnic State University, CSSE, San Luis Obispo, California, USA
{bhushan.chitre, jane.hayes}@uky.edu, dekhtyar@calpoly.edu

**Abstract.** **[Context and motivation]** Trace matrices are lynch pins for the development of mission- and safety-critical software systems and are useful for all software systems, yet automated methods for recovering trace links are far from perfect. This limitation makes the job of human analysts who must vet recovered trace links more difficult. **[Question/Problem]** Earlier studies suggested that certain analyst behaviors when performing trace recovery tasks lead to decreased accuracy of recovered trace relationships. We propose a three-step experimental study to: (a) determine if there really are behaviors that lead to errors of judgment for analysts, (b) enhance the requirements tracing software to curtail such behaviors, and (c) determine if curtailing such behaviors results in increased accuracy. **[Principal ideas/results]** We report on a preliminary study we undertook in which we modified the user interface of RETRO.NET to curtail two behaviors indicated by the earlier work. We report on observed results. **[Contributions]** We describe and discuss a major study of potentially unwanted analyst behaviors and present results of a preliminary study toward determining if curbing these behaviors with enhancements to tracing software leads to fewer human errors.

## 1 Introduction and Motivation

Automated tracing, generating or recovering the relationship between artifacts of the software development process, has been well researched over the past 15 years [4], but this automation doesn't come without inherent costs. One such cost is the need for human analysts to interact with the results of the automated methods. What we currently know about such interactions is that they tend to end disappointingly [1,2,6]. As long as we are using automated tracing methods for safety- and mission-critical systems, we must have humans vet the links. Therefore, we need to figure out how to make humans more accurate as they work with the results of automated methods. In prior studies we noticed some unwanted behaviors [1,2,6]. Can we curb them? Will curbing them yield fewer human errors?

A trace matrix is a collection of trace links, defined as "a specified association between a pair of artifacts, one comprising the source artifact and one comprising the target artifact." by the Center of Excellence for Software and System Traceability (COEST) [3]. A plethora of researchers have designed techniques for automatically or semi-automatically generating trace matrices, many discussed in a comprehensive survey by Borg [4]. Most of the focus in that work was on improving the quality of the candidate trace matrix, the matrix generated by a software method. While that work continues, recent work has segued into study of the analyst who works with the candidate matrix to generate the final trace matrix — the one that is used in application.

A typical trace tool, such as RETRO.NET used in this work [5], displays the candidate trace matrix and shows the list of source (high level) elements, and the list of candidate target (low level) elements that were automatically mapped to the source element. The texts of all elements can also be viewed. The key function of a tracing tool is to allow the analyst to vet individual candidate links.

Cuddeback et al. [1] and Dekhtyar et al. [2] studied the work of analysts with candidate trace matrices produced by automated software. The analysts were presented a candidate trace matrix and were asked to evaluate the individual links and correct any errors of omission or commission. The accuracy of candidate trace matrices varied from analyst to analyst — from high-accuracy matrices that contained few omitted links and few false positives to low-accuracy ones which contained many errors of both types. The studies found that analysts working with high accuracy candidate traces tended to decrease the accuracy — i.e., introduce false links into the matrix and remove true links, whereas the analysts who had low accuracy matrices tended to improve the accuracy significantly[1]. A follow-up study collected logs of analyst activity during the tracing process, and looked at behaviors that correlated with improved or decreased accuracy [6]. While that study did not have enough data points to allow for statistical significance of the results, the authors observed a number of analyst behaviors that tended to lead to errors of judgement. Specifically, two behaviors briefly described below were observed.

> **Long time to decide.** When analysts took unusually long (for their pace) time to decide whether a candidate link needed to be kept in the trace, *they tended to make an incorrect decision* [6].
>
> **Revisiting a link (backtracking).** When analysts revisited a link on which they already entered a decision and reversed that decision, *they tended to err* [6].

Our motivation for the continuing study of analyst behavior in tracing tasks comes from the key observations from the prior work [4,1,2,6]. On one hand, the lack of traceability as a byproduct of development in large software projects demonstrates a clear need for accurate automatic tracing methods [4]. At the same time, human analysts, when asked to curate automatically obtained traceability relations, make mistakes and decrease the overall accuracy of the trace [1,2]. We observe that one possible way to resolve this, and to improve the accuracy of curated trace relations is, potentially, to curb analyst behaviors that result in errors. In fact, psychologists studying human decision-making have observed that humans tend to operate in one of two decision-making systems — System 1 (S1) (or fast, instinctive thinking) or System 2 (S2) (slow, deliberate, logical thinking) [8]. The observed behaviors leading to decrease in accuracy belong to System 2. This motivates an additional research question expressed below.

---

[1] As reported earlier [2], the accuracy of the starting RTM affected the changes in precision, recall, and f2-measure, and the final precision in statistically significant ways, but did not affect final recall or final f2-measure in statistically significant ways.

## 2 Curbing Unwanted Analyst Behavior

The latter observation serves as the inspiration for our next step in the study of the behavior of human analysts. In this section we discuss the overall plan for the study, as well as the preliminary work we conducted.

**2.1 Research Preview**

The study we are planning to undertake consists of three key research questions.
1. **RQ1:** *Are there analyst behaviors that tend to reliably lead to analysts making errors, and where do these behaviors fall on the Kahneman's thinking system dichotomy [8]?* We hypothesize that such behaviors can be observed as statistically significant. We additionally conjecture that such behaviors would correspond to the decision-making System 2 [8].
2. **RQ2:** *What software enhancements for automated tracing tools can be designed and developed to curb the discovered unwanted behaviors?* We hypothesize that each unwanted behavior can be curbed via UI and workflow changes to the requirements tracing software.
3. *RQ3: Is there an improvement in the accuracy of final trace matrices constructed by the analysts using software with the implemented enhancements?* We hypothesize that the software enhancements will improve the accuracy (i.e., decrease the number of errors that analysts make in vetting candidate links and in discovery of omitted links).

The basic outline of the study is as follows.

**Discovery of analyst behaviors.** In the first stage we plan to replicate the tracing experiment of Kong et al. [6] in which we collected activity logs from a group of analysts performing a tracing task with a version of RETRO.NET enhanced with event logging. The original study included only a few data points, and did not allow the authors to observe any specific harmful behaviors with any degree of statistical rigor. Our intent is to collect significantly more data points (i.e., logs documenting analyst's work with a tracing tool on a tracing task), so that log analysis may reveal clear analyst behaviors that either tend to lead to errors, or tend to reliably improve accuracy, and provide more than just anecdotal evidence in support of such observations.

RETRO.NET logs information about individual analyst interactions with the software — keys pressed, elements selected, linking decisions made and confirmed, searches performed, etc. Each log record is keyed by a timestamp, making it easy to map analyst behavior, and in particular to map their correct and erroneous decisions along the time axis.

Initial replicated experiments were conducted in Spring 2017 and Fall 2017 quarters. We have been able to collect over 80 data points, and are currently in the process of analyzing the results to see if the prior observations [1,2] are confirmed. In the immediate future, we plan to replicate the analysis of Kong et al. [6] on the 80+ tracing logs we now have.

The first observed behaviors leading to errors belonged to Kahneman's System 2 (slow and deliberate) way of thinking. This leads us to ask the following question during the discovery process: *is RTM analysis a process that can be performed best within the System 1 (fast, intuitive) [8] of decision-making?* To answer this question, we can classify the observed harmful behaviors within the S1 — S2 dichotomy.

**Development of software enhancements.** Once we identify analyst behaviors that tend to lead to errors in link vetting, we plan to develop software-supported strategies for curbing such unwanted behaviors. For each behavior discovered, we will design one or more features to enhance RETRO.NET in a way that would reduce behavior incidence. We will explore the following approaches:

1. *Warnings.* This is a very basic approach: detect an unwanted behavior, and as soon as it is observed produce a warning within the tracing software suggesting that the analyst reconsider.
2. *Prohibitions.* This approach starts the same way as a warning with the detection of the unwanted behavior, but instead of simply producing a warning, the software will simply refuse to grant the analyst the ability to complete the unwanted behavior.
3. *Restructuring.* Certain unwanted behaviors may be eliminated or reduced if the way the analyst interacts with the tracing software is changed, and the use cases where such unwanted behaviors were observed are altered in significant ways. An example of a restructuring solution may be a change from allowing the analyst to review candidate links in arbitrary order to an interaction model where the analyst is shown each link once in a predefined order and is not allowed to revisit a link.

**Study of the impact.** We want to know the answers to two key questions:

1. *Do software enhancements designed to curb unwanted behaviors **actually curb** these behaviors?*
2. *Is the decrease in unwanted behaviors accompanied by a decrease in the number of errors analyst make?* (and thus by an increase in the accuracy of the trace relation).

To answer these questions we plan to conduct a second replication of the prior study [6], only this time we will use control and experimental groups of analysts. The control group will work with the standard version of the RETRO.NET tool, without any enhancements implemented in Stage 2 of the study. The experimental group will work with a version of RETRO.NET enhanced with specific solutions for curbing unwanted behavior. To test different ways of curbing the same behavior, we may need to conduct multiple rounds of such study.

### 2.2 Preliminary Study

To test the feasibility of our approach we conducted a preliminary study. We briefly describe the structure of the study and its results below.

**Unwanted analyst behaviors.** The study concentrated on the two analyst behaviors described in Section 1 (a) taking an *unusually* long amount of time to make a decision on a candidate link, and (b) revising an explicitly conveyed decision on a link. These were the two clearest behaviors observed previously [6] that tended to result in errors.

**Software enhancements.** We elected to start with very simple modifications to RETRO.NET. For each behavior, RETRO.NET was enhanced with code working in the background designed to detect it, and with UI elements that would produce a warning message to the analyst when the behavior was discovered. Specifically, the enhanced RETRO.NET, upon detecting either of the two behaviors, displayed a pop-up window informing the user that their behavior could lead to an error. In the case of the user trying to revisit a decision, the user is given an option to backtrack. In both cases, the user can also dismiss the prompt and simply continue with their action. In making decisions about the enhancements of RETRO.NET we tried to make the changes simple and non-prohibitive. We understand that UI design principles suggest that pop-up messages that disrupt the flow of user interaction with the software may reduce productivity and decrease user satisfaction with the software and its UI. At the same time, we wanted the warnings in our first experiment to be "blatant," easy to see, and hard to miss. We took the risk of implementing the warnings via the pop-up message UI elements fully realizing that we may be sacrificing some user satisfaction with the software.

**The study.** A total of 14 subjects participated in a preliminary study conducted in Spring of 2017 at the University of Kentucky. Five (5) subjects were in the control group and worked with non-enhanced RETRO.NET. Nine (9) subjects were in the experimental group and worked with the RETRO.NET version enhanced with backtracking and taking-too-long warnings[2]. Each subject received a brief training session on their version of RETRO.NET using the same toy dataset. Later, they were presented with the ChangeStyle dataset [1,2] to trace. All subjects started with the same initial candidate trace matrix. We measured the *precision, recall, f2-measure,* and *lag* [7] of the resulting trace matrix the subjects submitted and the *time* it took them to complete the work. The results of the preliminary study are shown below.

### 2.3 Preliminary Study Results

In our preliminary study, the experimental group showed higher mean precision (15.6% vs. 8.3%), higher mean recall (96% vs. 77.6%), and higher mean f2-measure (0.329 vs 0.262), as well as better (lower) lag (1.85 vs. 2.55) for the submitted traces. Only two mean values were better for the control group: the mean time (75 minutes versus 82) and the change in true positives was higher (1.6 versus 1.222). This could be explained by the extra prompts that were shown to the user: (a) that had to at least be dismissed, and (b) that had to be at most obeyed.

---

[2] Originally, the control and the experimental groups were of the same size, but we had a significantly larger number of non-completions in the control group.

## 3    Discussion and Conclusions

The preliminary study tentatively indicates that basic prompts (discussed in Section 2.1 as warnings) may suffice to move analysts away from undesired behaviors without having to resort to more restrictive measures, but at the expense of time taken to perform tracing. The main, and very useful, outcome of the preliminary study is a list of items that we must add to our future study: collect the number of times that prompts appear, collect the amount of time that an analyst takes when dismissing and reacting to the prompt, track the action taken by the analyst after a prompt, track the number of false positives (true negatives, false positives, and false negatives) added and removed, and potentially track each individual true positive link displayed by RETRO.NET to learn its final disposition.

As mentioned in Section 2.1, we envision a three stage approach to investigating our main research question: can we help analysts vet trace matrices? For the first phase of the study, discovery of the analyst behaviors leading to errors, we plan to undertake studies (and have already undertaken some of them) using a software tracing tool in order to discover what behaviors analysts exhibit when tracing. We posit that we will discover good behaviors (those that lead to improved trace matrices) as well as unwanted behaviors - those that lead to errors. Our early work discussed above is a first step toward addressing the second of the three phases: enhance tracing software to curtail unwanted behaviors and learn whether or not the software enhancements do indeed curtail them. For phase three, we plan to undertake a study similar to that of our preliminary study, but with a wider scope. We plan to collect richer data from significantly larger control and experimental groups. We also envision undertaking a statistical study of our data, as we will have sufficient data points to permit such analysis. It is our hope that these three stages of our study will contribute to our field and more importantly to software tracing tools put in the hands of practitioners so that analyst tracing work won't end in disappointment, but rather in effective and efficient use of the analysts' time.

ACKNOWLEDGMENT

We thank Dr. Dan Berry for insightful comments and suggestions on prior versions that resulted in a greatly improved paper. We thank all participants from upper division software engineering classes who took their time to participate in our study. We thank NASA and NSF as prior grants funded the development of RETRO.NET. We thank Jody Larsen, the developer of RETRO.NET. We thank NSF for partially funding this work under grants CCF-1511117 and CNS- 1642134.